\documentclass[twocolumn, pra, showpacs,superscriptaddress]{revtex4}
\usepackage{amssymb}
\usepackage{graphicx}
\usepackage{dcolumn}
\usepackage{bm}
\usepackage{amsmath}

\begin{document}

\title{Improvement of the matching of the exact solution and variational
approaches in an interacting two-fermion system}
\author{Yanxia Liu}
\affiliation{Department of Physics and Institute of Theoretical Physics, Shanxi
University, Taiyuan 030006, P. R. China}
\author{Jun Ye}
\affiliation{Department of Physics and Institute of Theoretical Physics, Shanxi
University, Taiyuan 030006, P. R. China}
\author{Yuanyuan Li}
\affiliation{Department of Physics and Institute of Theoretical Physics, Shanxi
University, Taiyuan 030006, P. R. China}
\author{Yunbo Zhang}
\email{ybzhang@sxu.edu.cn}
\affiliation{Department of Physics and Institute of Theoretical Physics, Shanxi
University, Taiyuan 030006, P. R. China}

\begin{abstract}
A more reasonable trial ground state wave function is constructed for the
relative motion of an interacting two-fermion system in a 1D harmonic
potential. At the boundaries both the wave function and its first derivative
are continuous and the quasi-momentum is determined by a more practical
constraint condition which associates two variational parameters. The upper
bound of the ground state energy is obtained by applying the variational
principle to the expectation value of the Hamiltonian of relative motion on
the trial wave function. The resulted energy and wave function show better
agreement with the analytical solution than the original proposal.
\end{abstract}

\maketitle

\section{Introduction}

Tremendous progress in cooling, trapping and manipulating ultracold samples
makes it possible to experimentally study few-body phenomena in trapped
atomic and molecular systems with precise control \cite{DBlume}. The study
of trapped few-atom systems provides a bridge between small and large
systems and the treatment of such mesoscopic systems is complementary to the
interpretation derived within effective many-body frameworks. Systems made
of just two interacting atoms are especially relevant due to its role as
building blocks of many-body strongly correlated states. Two-body system
with tunable interactions has been realized in the Heidelberg experiment
\cite{Zurn} using two fermionic $^{6}$Li atoms in the ground state of a
potential created by an optical dipole trap. In this experiment, the
interaction energy of two distinguishable fermions as a function of the
interaction strength has been measured showing the experimental capability
to simulate strongly correlated few-body quantum systems. An exceptional
property of the two-body interacting system is that there exists analytical
solution for arbitrary values of the interparticle interaction \cite%
{TBusch,ZIdziaszek}. A theory for the tunneling of one atom out of a trap
containing two interacting cold atoms is developed by introducing the
quasi-particle wave function \cite{Rontani}. Unfortunately, no theoretical
solution is available for more complex systems even if one more atom is
included.

It is thus of interest to seek for some approximation methods which may
potentially be generalized to more than two particles. Ref. \cite{DRubeni}
provides such an example, which calculated the ground-state energy of the
relative motion of a system of two fermions with spin up and spin down
interacting via a delta-function potential in a one-dimensional (1D)
harmonic trap. The authors divided the relative coordinates space into three
regions and established a trial ground state wave function according to the
Bethe-ansatz solution for a system with delta-function interaction and the
ground state wave function of harmonic potential. They directly applied the
periodic boundary conditions to obtain the Bethe ansatz equation, which was
taken as the constraint condition for quasi-momentum. By analyzing solution
of the Bethe ansatz equation, we know that the possible values for the
quasi-momenta $k$ depend on the value of $c$ \cite{MTakahashi}. For the
repulsive case ($c>0$), only real values of $k$ are ground-state solutions
of the Bethe ansatz equation. For the attractive case($c<0$), only the
purely imaginary numbers $k$ satisfy for the ground state. For both cases
they found the numerical minimization of expectation value of the ground
state energy by introducing two variational parameters. This work provides a
starting point for the investigation of more complex few-body systems where
no exact theoretical solution is available.

In this paper we improve the trial ground state wave function proposed in
\cite{DRubeni} from two aspects. Firstly not only wave function but also its
derivative are continuous at the boundaries, which avoids introducing the
discontinuity of the first derivative of the wave function for regular
potential. Secondly instead of using Bethe ansatz equation as the constraint
condition, we find a more practical constraint condition for the
quasi-momentum which incorporates the two variational parameters and the
contact interaction strength. There, the solutions of the constraint
condition no longer depend on the sign of $c$.

The paper is organized as follows. In Sec. II, we introduce the modified
trial ground state wave function for the system of two interacting fermionic
atoms. At the boundaries both the wave function and its first derivative are
continuous and we obtain a similar constraint condition for the two
variational parameters. Then the variational principle is applied to the
expectation value of the Hamiltonian of relative motion on the conjectured
wave function in Sec. III. In Sec. IV Our results are compared with the
analytical result obtained in \cite{TBusch,ZIdziaszek} and the approximate
result obtained in \cite{DRubeni}. In addition the density distributions for
various interaction constants also show good agreement with the analytical
solution. We conclude our results in Sec. V.

\section{Trial Wave Function}

We shall consider the same system of two interacting fermionic atoms with
mass $m$ loaded in a harmonic trap of frequency $\omega $ as in Ref. \cite%
{DRubeni}. The Hamiltonian can be formulated as
\begin{equation}
H=-\frac{\hbar ^{2}}{2m}\frac{\partial ^{2}}{\partial x_{1}^{2}}-\frac{\hbar
^{2}}{2m}\frac{\partial ^{2}}{\partial x_{2}^{2}}+V_{A}\left(
x_{1},x_{2}\right) +V_{I}\left( x_{1},x_{2}\right)  \label{h}
\end{equation}%
where $V_{A}\left( x_{1},x_{2}\right) $ is the trapping potential for atoms
located at $x_{1}$ and $x_{2}$
\begin{equation}
V_{A}\left( x_{1},x_{2}\right) =\frac{1}{2}m\omega ^{2}x_{1}^{2}+\frac{1}{2}%
m\omega ^{2}x_{2}^{2}.
\end{equation}%
The interaction potential $V_{I}$ is assumed to be \emph{s}-wave contact
potential represented by a delta-function
\begin{equation}
V_{I}\left( x_{1},x_{2}\right) =2c\delta \left( x_{2}-x_{1}\right)
\end{equation}%
where the interaction strength $c$ can be tuned via magnetic Feshbach
resonance from $-\infty $ to $+\infty $.

Introducing the center-of-mass and relative coordinates $%
x=x_{1}-x_{2},X=(x_{1}+x_{2})/2$ allows us to decompose the Hamiltonian (\ref%
{h}) into the center-of-mass part
\begin{equation}
H_{CM}=-\frac{\hbar ^{2}}{2M}\frac{\partial ^{2}}{\partial X^{2}}+\frac{1}{2}%
M\omega ^{2}X^{2}
\end{equation}%
and relative motion part%
\begin{equation}
H_{rel}=-\frac{\hbar ^{2}}{2\mu }\frac{\partial ^{2}}{\partial x^{2}}%
+2c\delta \left( x\right) +\frac{1}{2}\mu \omega ^{2}x^{2}.
\end{equation}%
The Hamiltonian of the center-of-mass is nothing but a simple harmonic
oscillator with total mass $M=2m$, while the relative part describes an
oscillator with reduced mass $\mu =m/2$ and a delta potential well/barrier
in the trap center. The latter has been exactly solved in Ref. \cite{TBusch}.

In Ref. \cite{DRubeni} the authors built a variational trial wave function $%
\psi (x,\{\alpha ,L\})$ based on the Bethe ansatz solution of two particles
and it was shown that applying the same principle for a large number of
fermions is possible. The parameters $\alpha $ and $L$ controls the decay of
the trial function outside the trap and the region where the decay occurs.
Minimizing the expectation value of the Hamiltonian $H_{rel}$ on the
conjectured wave function $\psi (x,\{\alpha ,L\})$ by setting the
derivatives with respect to these two parameters zero gives the upper bound
of the ground state energy of the system
\begin{equation}
E_{GS}\leq \frac{\left\langle \psi \right\vert H_{rel}\left\vert \psi
\right\rangle }{\left\langle \psi |\psi \right\rangle }  \label{3rt}
\end{equation}%
The trial wave function $\psi $ is continuous in Ref. \cite{DRubeni},
however, the first derivatives of the wave function develop an unphysical
discontinuity at $x=\pm L$. We know from basic quantum mechanics that in
matching the wave functions in different regions the wave function $\psi $
is always continuous and the first derivative $d\psi /dx$ is continuous
except at points where the potential is infinite, i.e. $\delta $ potential.
In our problem there do not exist such $\delta $ potential points at $x=\pm
L $ hence discontinuity of the first derivative there should be excluded. As
we shall see below a simple improvement of the trial wave function will help
us get rid of this dilemma and the resulted ground state energy matches the
analytical result even better.

\begin{figure}[tbp]
\includegraphics[width=0.5\textwidth]{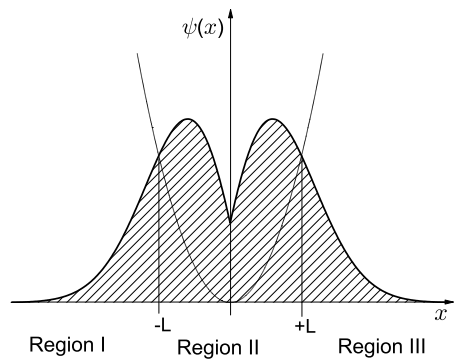}
\caption{The potential and the trial ground state wave function $\Psi (x)$
in the relative coordinates system of interacting two-fermion system. Both
the wave functions and its derivatives are continuous at the boundary $x=\pm
L$.}
\label{fig1}
\end{figure}

In the following we introduce the variational ansatz and the modification of
the trial ground state wave function. As shown in Fig. \ref{fig1}, the wave
function is continuous and it's derivative is also continuous, which is much
closer to the real wave function.

How do we build the trial ground state wave function for the two body
problem of Hamiltonian $H_{rel}$? The space is divided into three regions by
the parameter $L$. The wave functions in the regions I and III are chosen as
eigenstates of 1D harmonic oscillator center at $x=0$, instead of at $x=\pm
L $ as done in Ref. \cite{DRubeni}. $\psi _{I}$ and $\psi _{III}$ have the
form of the Gaussian with the parameter $\alpha $. In the regions II, i.e.
in the vicinity of the $\delta $ potential center, the contact interaction
term plays the dominant role and the harmonic potential can be neglected.
The wave function takes the form of\ two distinct fermions with a contact
interaction. Formally we assumes the trial ground state wave function as the
following configuration%
\begin{equation}
\psi =\left\{
\begin{array}{ll}
\psi _{I}=Ae^{-\alpha x^{2}} & -\infty <x<-L \\
\psi _{II}=\left( Be^{-ikx}+Ce^{ikx}\right) \Theta \left( x\right) &  \\
\qquad +\left( De^{ikx}+Ee^{-ikx}\right) \Theta \left( -x\right) & -L<x<L \\
\psi _{III}=Fe^{-\alpha x^{2}} & L<x<\infty%
\end{array}%
\right.
\end{equation}%
where $\Theta $ is the Heaviside step function%
\begin{equation}
\Theta (x)=\left\{
\begin{array}{cc}
1, & x>0 \\
0, & x<0%
\end{array}%
\right.
\end{equation}

The remaining work is to match the wave functions in different regions
through the boundary conditions. In region II, the continuity of the wave
function $\psi _{II}$ at $x=0$ gives the following coefficient relation
\begin{equation}
B+C=D+E.  \label{a11}
\end{equation}%
By integrating the eigenequation $\hat{H}\psi _{II}=E\psi _{II}$ from the
negative infinitesimal to the positive infinitesimal, we do find a jump of
the first derivative of wave function at $x=0$
\begin{equation}
ik(-B+C-D+E)=\frac{2\mu }{\hbar ^{2}}\times 2c(B+C).  \label{a13}
\end{equation}%
Because the potential is axially symmetric, the ground state wave function
must be spatially symmetric and has zero nodes, so%
\begin{equation}
B=D,C=E,A=F.  \label{a15}
\end{equation}%
According to above three equations, we get%
\begin{equation}
B=\frac{ik-\frac{2\mu c}{\hbar ^{2}}}{ik+\frac{2\mu c}{\hbar ^{2}}}C.
\label{a14}
\end{equation}%
At $x=-L$ or $x=L$, the wave function is continuous, which means%
\begin{equation}
Ae^{-\alpha L^{2}}=C\left( \frac{ik-\frac{2\mu c}{\hbar ^{2}}}{ik+\frac{2\mu
c}{\hbar ^{2}}}e^{-ikL}+e^{ikL}\right)  \label{c1}
\end{equation}%
and the first derivative of $\psi $ is continuous too, we get%
\begin{equation}
2\alpha Le^{-\alpha L^{2}}A=Cik\left( \frac{ik-\frac{2\mu c}{\hbar ^{2}}}{ik+%
\frac{2\mu c}{\hbar ^{2}}}e^{-ikL}-e^{ikL}\right)  \label{c2}
\end{equation}%
The combination of (\ref{c1}) and (\ref{c2}) leads us to a constraint
condition%
\begin{equation}
e^{i2kL}=\frac{k+i\frac{2\mu }{\hbar ^{2}}c}{k-i\frac{2\mu }{\hbar ^{2}}c}%
\frac{k+i2L\alpha }{k-i2L\alpha }.  \label{2as}
\end{equation}
This equation plays similar role as that of Bethe ansatz equation. It is,
however, irrelevant with the boundary conditions (periodic or open) for the
two-fermion system which should always be chosen before one deals with the
exactly solvable models. In this way we avoid artificially imposing the
periodic boundary condition onto the system. And we find no obstacles to
generalize this constraint condition to $N$ fermions following a similar
scheme as in the Appendix of Ref. \cite{DRubeni}.

On the basis of above restrictions to the coefficients, we get the following
trial ground state wave function subject to normalization%
\begin{equation}
\psi =\left\{
\begin{array}{ll}
\psi _{I}=\psi _{II}(-L)e^{-\alpha x^{2}}e^{\alpha L^{2}} & -\infty <x<-L \\
\psi _{II}=\cos (\theta -kx)\Theta \left( x\right) &  \\
\qquad +\cos (\theta +kx)\Theta \left( -x\right) & -L<x<L \\
\psi _{III}=\psi _{II}(L)e^{-\alpha x^{2}}e^{\alpha L^{2}} & L<x<\infty%
\end{array}%
\right.  \label{wavefunction}
\end{equation}%
where $\theta $ satisfies the following relation%
\begin{equation}
\theta =\arctan \frac{2\mu c}{k\hbar ^{2}}.
\end{equation}%
This trial ground state wave function, composed of pure elementary
functions, offers convenient analysis due to the continuity of both its
first derivative and itself. We schematically show the configuration of the
potential and the wave functions in Figure \ref{fig1}, which join each other
smoothly at the boundaries for all three regions.

\begin{figure}[tbp]
\includegraphics[width=0.5\textwidth]{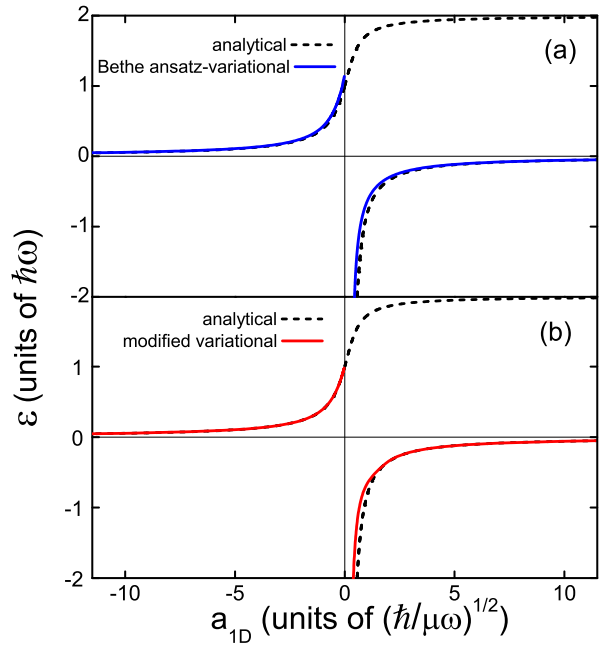}
\caption{(Color online)The ground state energies of the relative motion for
two distinct fermions interacting via s-wave pseudopotential confined in an
axially symmetric harmonic trap with angular frequency $\protect\omega $.
The black dotted lines indicate energy levels of analytical results. (a) The
blue curves represent the energy by using Bethe ansatz equation as
constraint condition. (b) The red curves represent the energy by using
constraint condition of the continuity of wave function derivative.}
\label{fig2}
\end{figure}

\section{Variational Principle}

We choose $\alpha $ and $L$ as our variational parameters and aim to find
the least possible value of the ground state energy on the trial wave
function for different interaction strength $c$ subject to the constraint
condition (\ref{2as}). Contrary to the original proposal (eq. (3.2) in \cite%
{DRubeni}), in our scheme, both variational parameters $\alpha $ and $L$
enter into the constraint condition, which determines the quasi momentum $k$%
. To apply the variational principle, we need to compute the normalization
factor of the wave function, which yields%
\begin{eqnarray}
\left\langle \psi |\psi \right\rangle &=&\sqrt{\frac{\pi }{8\alpha }}\frac{%
1+\cos (2kL-2\theta )}{e^{-2\alpha L^{2}}}(1-\text{erf}(\sqrt{2\alpha }L))
\notag \\
&&+\frac{1}{2k}\left( \sin \left( 2kL-2\theta \right) +\sin 2\theta
+2kL\right)  \label{zz1}
\end{eqnarray}%
where the error function $\text{erf}(x)$ is defined as%
\begin{equation}
\text{erf}(x)=\frac{2}{\sqrt{\pi }}\int_{0}^{x}e^{t^{2}}dt.
\end{equation}%
Moreover, the expectation value of $H_{rel}$ on the ground state (\ref%
{wavefunction}) is calculated as
\begin{widetext}
\begin{eqnarray}
\left\langle \psi \right\vert H_{rel}\left\vert \psi \right\rangle  &=&\left[
1+\cos (2kL-2\theta )\right] \left( 4\alpha ^{2}\hbar ^{2}+\mu ^{2}\omega
^{2}\right) \left( \sqrt{\frac{\pi }{2\alpha }}\frac{1-\text{erf}(2\alpha L))%
}{16\mu \alpha e^{-2\alpha L^{2}}}+\frac{L}{8\mu \alpha }\right)   \notag \\
&&-\frac{\hbar ^{2}k}{4\mu }\left[ \sin \left( 2kL-2\theta \right) +\sin
\left( 2\theta \right) \right] +\frac{\hbar ^{2}k^{2}L}{2\mu }+\frac{\mu
\omega ^{2}L^{3}}{6}+c\left( 1+\cos 2\theta \right)   \notag \\
&&+\frac{\mu \omega ^{2}}{8k^{3}}\left[ \left( 2k^{2}L^{2}-1\right) \sin
\left( 2kL-2\theta \right) +2kL\cos \left( 2kL-2\theta \right) -\sin 2\theta %
\right]   \label{zz2}
\end{eqnarray}
\end{widetext}

It is interesting to consider first the limiting case of $L\rightarrow
0,c\rightarrow 0$, that is, a simple harmonic oscillator. Accordingly the
upper bound for the ground state energy (\ref{3rt}) reduces to%
\begin{equation}
\lim_{L,c\rightarrow 0}\frac{\left\langle \psi \right\vert H_{rel}\left\vert
\psi \right\rangle }{\left\langle \psi |\psi \right\rangle }=\frac{\hbar
^{2}\alpha }{2\mu }+\frac{\mu \omega ^{2}}{8\alpha }
\end{equation}%
The minimum value of the energy can be found by setting\ the first
derivative with respect to $\alpha $ to zero. We find the extreme value of $%
\alpha $, which indicates the decay of the trial function, to be $\alpha
^{\ast }=\mu \omega /2\hbar $, such that%
\begin{equation}
\lim_{L,c\rightarrow 0}\frac{\left\langle \psi \right\vert H_{rel}\left\vert
\psi \right\rangle }{\left\langle \psi |\psi \right\rangle }|_{\alpha
=\alpha ^{\ast }}=\frac{1}{2}\hbar \omega
\end{equation}%
In this case we recover precisely the ground state energy of the simple
harmonic oscillator.

\section{Comparison}

For the model system the constraint condition (\ref{2as}) for the trial wave
function reduces to%
\begin{equation}
\alpha =\frac{k^{2}\hbar ^{2}\tan \left( kL\right) -2c\mu k}{2\hbar
^{2}Lk+4\mu cL\tan \left( kL\right) }.  \label{q1}
\end{equation}%
Putting this back into (\ref{3rt}), we sweep over all values of $L$ and $k$
for each assigned $c$ and minimize the expression (\ref{3rt}) numerically.
We call the extreme parameters $L^{\ast }$ and $k^{\ast }$ when the
minimization point has been reached. In this way we determine the ground
state energy of the two-fermion system as a function of the coupling $c$ via
the variational principle, for both repulsive and attractive interaction.
The result is depicted in Fig. \ref{fig2} in terms of the physical variables
$\varepsilon $ and $a_{1D}$, which are related to the energy $E_{GS}$ and
interaction strength $c$ through%
\begin{eqnarray}
E_{GS} &=&\left( \varepsilon +\frac{1}{2}\right) \hbar \omega ,  \notag \\
a_{1D} &=&-\frac{1}{2c}\sqrt{\frac{\hbar ^{3}\omega }{\mu }}
\end{eqnarray}%
and the value of$\ \varepsilon $ is determined by the follow implicit
equation%
\begin{equation}
2a_{1D}=\frac{\Gamma (-\frac{\varepsilon }{2})}{\Gamma (-\frac{\varepsilon }{%
2}+\frac{1}{2})}
\end{equation}%
where $\Gamma (x)$ is the complete gamma function. We notice that by
combining the variational principle and continuity of the wave function
derivatives a better agreement is achieved between this analytical result
(black dotted line in Fig. \ref{fig2}) and our result (red solid line in Fig. %
\ref{fig2}b) than that\ obtained with the Bethe ansatz equation as
constraint condition (blue solid line in Fig. \ref{fig2}a). In the repulsive case
the modified variational curve lies exactly on the top of analytical one and the
effective matching
range of the approximation in the attractive case is also much wider than
the original proposal.

\begin{figure}[tbp]
\includegraphics[width=0.50\textwidth]{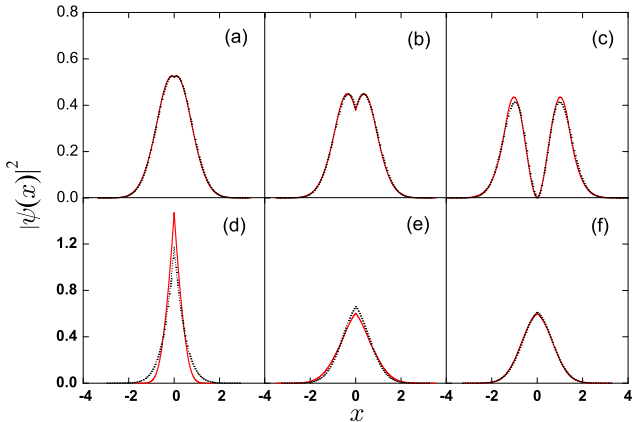}
\caption{(Color online)The normalized probability density in the relative
coordinates of two interacting fermions. Here the inter-particle interactions
are in units of $\protect\sqrt{\protect\mu \diagup \hbar ^{3}\protect\omega }$
and the 1D scattering lengths are in units of $\sqrt{\hbar/\mu \omega}$:
(a) $c=0.05, a_{1D}=-10$; (b) $c=0.25, a_{1D}=-2$; (c) $c=20, a_{1D}=-0.025$%
; (d) $c=-0.5, a_{1D}=1$; (e) $c=-0.1, a_{1D}=5$; (f) $c=-0.05, a_{1D}=10$. The
red solid lines show the analytical solutions of the probability density
while the black dotted lines show the variational results with modified trial
wave functions.}
\label{full}
\end{figure}

Inserting the parameters$\ L^{\ast }$ and $k^{\ast }$ back into the equation
(\ref{q1}), we get the corresponding$\ \alpha ^{\ast }$. The trial wave
function (\ref{wavefunction}) is determined by these parameters$\ L^{\ast }$,%
$\ \alpha ^{\ast }$ and $k^{\ast }$ for each value of $c$. The probability
densities of the trial ground state wave function in the relative coordinate
are illustrated in Fig. \ref{full} over the whole range of $c$. It reveals
that for repulsively interacting system
our variational wave functions near perfectly match the analytical
ground state wave functions of the relative motion Hamiltonian $H_{rel}$
(see Fig. \ref{full}a-\ref{full}c),
which is \cite{TBusch}%
\begin{equation}
\varphi _{0}\left( x\right) =A_{0}U\left( -\frac{\varepsilon }{2}%
,0.5,2\alpha x^{2}\right) \exp \left( -\mu \omega x^{2}/2\hbar \right).
\end{equation}%
The extension to weakly attractive case is possible (Fig \ref{full}e and \ref{full}f),
but not too far. If the interaction
is negative, the ground state is McGuire¡¯s cluster state
\cite{McGuire,Muth}, which is a bound state and would decay quickly
via molecular channels. We see that whereas our variational theory yields
surprisingly accurate energies, the wave functions are notoriously poor
for finite attraction (Fig. \ref{full}d).

\section{Conclusion}

In conclusion, we have found a much better trail wave function that
approximates the ground state energy of two interaction fermions in a 1D
harmonic trap via variational principle with the constraint conditions of
the continuity of both wave function and its derivative. This provides a
modified scheme which enables us to get rid of the idealized\ periodic
boundary condition in the central region $-L<x<L$. Instead we connect the
wave functions and derivatives at $x=\pm L$ by simple quantum mechanics
rules. A constraint condition is obtained to determine the quasi momentum,
under which the variational principle is applied to minimize the ground
state energy. This offers a new approach to solve the few-body problem and
the better agreement between our results and the analytical solution shows
that it is convincing to extend our method to many-body systems, which
arouse experimental interest \cite{FSerwane}.

\begin{acknowledgments}
This work is supported by NSF of China under Grant No. 11104171, No.
11074153 and No. 11234008, 973 Program under Grant No. 2010CB923103 and No.
2011CB921601.
\end{acknowledgments}

\end{document}